\documentclass[fleqn,usenatbib]{mnras}

\usepackage{newtxtext,newtxmath}

\usepackage[T1]{fontenc}
\usepackage{ae,aecompl}

\usepackage{graphicx}	
\usepackage[export]{adjustbox}
\usepackage{amsmath}	

\usepackage{color}

\newcommand{\msun}{\,$M_{\odot}$}
\newcommand{\ergs}{\,erg\,s$^{-1}$}
\newcommand{\kms}{\,km\,s$^{-1}$}

\newcommand{\cmq}{\,cm$^{-3}$}

\title[SNR candidate G116.6-26.1]{SRG/eROSITA discovery of a large circular SNR candidate G116.6-26.1: SN~Ia explosion probing the gas of the Milky Way halo?}

\author[Churazov et al.]{E.M.~Churazov,$^{1,2}$ I.I.~Khabibullin,$^{1,2}$ A.M.~Bykov,$^3$ N.N.~Chugai,$^4$ R.A.~Sunyaev,$^{1,2}$ and I.I.~Zinchenko$^{5}$
\\
\\
$^1$~Space Research Institute (IKI), Profsoyuznaya 84/32, Moscow 117997, Russia \\
$^2$~Max Planck Institute for Astrophysics, Karl-Schwarzschild-Str. 1, D-85741 Garching, Germany  \\
$^3$~Ioffe Institute, 26 Politekhnicheskaya str., St. Petersburg 194021, Russia \\
$^4$~Institute of Astronomy, Russian Academy of Sciences, 48 Pyatnitskaya str., Moscow 119017, Russia  \\
$^5$~ Institute of Applied Physics of the Russian Academy of Sciences, 46 Ul'yanov~str., Nizhny Novgorod 603950, Russia. \\
}

\date{Accepted XXX. Received YYY; in original form ZZZ}

\begin{document}
\label{firstpage}
\pagerange{\pageref{firstpage}--\pageref{lastpage}}
\maketitle

\begin{abstract}
We report a discovery of a new X-ray-selected supernova remnant (SNR) candidate SRGe~J0023+3625 = G116.6-26.1 found in the  \textit{SRG}/eROSITA all-sky survey. The source features a large angular extent ($\sim 4$ deg in diameter), nearly circular shape and X-ray spectrum dominated by emission lines of helium- and hydrogen-like oxygen. It lacks bright counterparts of similar extent at other wavelengths which could be unequivocally associated with it. Given the relatively high Galactic latitude of the source, $b\approx-26$ deg, we interpret these observational properties as an indication of the off-disk location of this SNR candidate. Namely, we propose that this object originated from a Type Ia supernova which exploded some 40 000 yr ago in the low density ($\sim 10^{-3}\,{\rm cm^{-3}}$) and hot ($\sim (1-2)\times10^6\,{\rm K}$) gas of the Milky Way halo at a distance of $\sim 3\,{\rm kpc}$ from the Sun.  The low density of the halo gas implies that the cooling and collisional ionization equilibrium (CEI) timescales downstream of the forward shock are much longer than the age of the SNR. This results in a relatively soft spectrum, reflecting pre-shock ionization state of the gas, and strong boost in the plasma emissivity (compared to CEI) due to enhanced collisional excitation through the increased electron temperature. If confirmed, such a rare object would provide us with a unique "in situ" probe of physical conditions (density, temperature and metallicity) near the interface between the Milky Way's disk and the halo.  
\end{abstract}


\begin{keywords}
ISM: supernova remnants -- radiation mechanisms: thermal -- X-rays: general -- Galaxy: halo 
\end{keywords}



\section{Introduction}

~~~~Remnants of supernova explosions (SNRs) provide us important information on the structure of the exploding star and, also, on the properties of the surrounding medium \citep[e.g.][]{1974ApJ...188..501C,1977ApJ...218..148M,2008ARA&A..46...89R}. Violent interaction between ejecta and the surrounding medium gives rise to the shock waves, gas heating and compression, accompanied by efficient particle acceleration to very high energies. As a result, these objects are often very bright over the entire electromagnetic spectrum, from low-frequency radio waves all the way to the TeV gamma-rays. Some 300 Galactic SNRs are listed in the current catalogs \citep[e.g.][]{2019JApA...40...36G}. Core-collapse SNRs are more numerous; they are associated with massive stars and, therefore, are strongly concentrated towards the Galactic Plane. Less frequent thermonuclear SNRs reflect the distribution of low-mass stars and, therefore, could be found at larger Galactic latitudes. 

While the majority of SNRs are found in the radio band, X-ray surveys remain an important source of new interesting candidates.  With the new \textit{SRG}/eROSITA X-ray survey covering the full sky one could expect that more SNR candidates will be found, in particular those featuring large angular size and relatively low surface brightness of the X-ray emission \citep[as exemplified by the recent discovery of the Hoinga  supernova remnant,][]{2021A&A...648A..30B}.

The \textit{SRG} X-ray observatory \citep{2021arXiv210413267S} with the two wide-angle grazing-incidence X-ray telescopes, eROSITA \citep[0.3-10 keV,][]{2021A&A...647A...1P} and Mikhail Pavlinsky ART-XC telescope \citep[4-30 keV,][]{2021arXiv210312479P}, was launched on July 13, 2019, and started its all-sky X-ray survey on December 13, 2019, upon completion of the commissioning, calibration, and performance verification observations. By mid June 2021, three consecutive scans of the entire sky have been acquired, resulting in the accumulated exposure time from $\sim700$ s at the ecliptic equator to $\sim$130 ks at the ecliptic poles. Thanks to the exquisite sensitivity at soft X-ray energies and stable particle background, \textit{SRG}/eROSITA allows mapping and accurate characterization of such large diffuse structures like eROSITA bubbles \citep{2020Natur.588..227P} or Cygnus star formation region, as well as bright supernova remnants like Cygnus Loop or Vela. 

Here we report discovery of a much fainter object,   SRGe~J0023+3625 = G116.6-26.1, which features large angular extent ($\sim 4$ deg in diameter), nearly circular shape and soft spectrum of the X-ray emission, dominated by emission lines of helium- and hydrogen-like oxygen. Taking into account its relatively large Galactic latitude, $b\approx-26$ deg, absence of bright counterparts at other wavelengths and absorption at soft energies  consistent with line-of-sight integral in that direction, we propose that this object might be a relatively old ($\sim$ 40 000 yrs) remnant of an SN~Ia explosion happened in the Galactic halo $\sim3$ kpc away from the Sun,i.e. $\sim$ 1 kpc out of the Galactic disk. In such a case, this object offers us a very rare opportunity to reveal properties of the hot gas near the Galactic disk-halo interface and directly observe non-equilibrium dynamics of plasma heating and ionization.
Future sensitive observations at radio, IR/optical and UV wavelengths should confirm or reject this scenario and complement the X-ray picture with more details regarding magnetic fields, particle acceleration and multiphase nature of the halo gas.



\section{Observations and Data}
\label{sec:obs}

At the moment of writing, the relevant part of the sky was covered by 3 consecutive \textit{SRG} all-sky surveys. To produce the calibrated event lists, the raw data were reduced using the \texttt{eSASS} software. Based on the analysis of the light curves, the data were cleaned from obvious artifacts and background flares. After the data cleaning, the accumulated exposure is fairly uniform over the field of interest ($\sim 800\,{\rm s}$ per point). The detector intrinsic background was estimated using the data from the all-sky survey collected during time intervals when the filter wheels of the telescope modules were in the "CLOSED" position, i.e. when the telescope detectors were not exposed to the X-ray emission of astrophysical origin \citep[see, e.g.][]{2021arXiv210413267S,2020SPIE11444E..1OF}. 

\subsection{X-ray imaging and spectral analysis}

The 0.5-0.7 keV image (detector intrinsic background subtracted and exposure corrected) in shown in the left panel of Fig.~\ref{fig:image}. The image was smoothed with the kernel corresponding to the survey PSF \citep[see Appendix B in][]{2020arXiv201211627C} and adaptively re-binned to suppress the noise and enhance visibility of faint diffuse emission. The right panel shows the same image after subtracting  compact sources ($F_X\gtrsim10^{-13}\,{\rm erg\,s^{-1}\,cm^{-2}}$) and smoothing with a broad Gaussian filter ($\sigma=4'$). Apart from large scale variations caused by the foreground absorption (as highlighted by anti-correlation of the soft X-ray surface brightness with the dust emission at 100 $\mu$m shown by black contours), a nearly circular structure with the radius $R\approx 1.95^\circ$ is clearly seen (see Table~\ref{tab:obs}) in both images.  The structure is mildly edge brightened with a (marginal) evidence of the brightening inside inner $20'$ as illustrated in Fig.~\ref{fig:radial}. 


\begin{figure*}
\centering
\includegraphics[angle=0,bb=110 200 580 670,width=0.99\columnwidth]{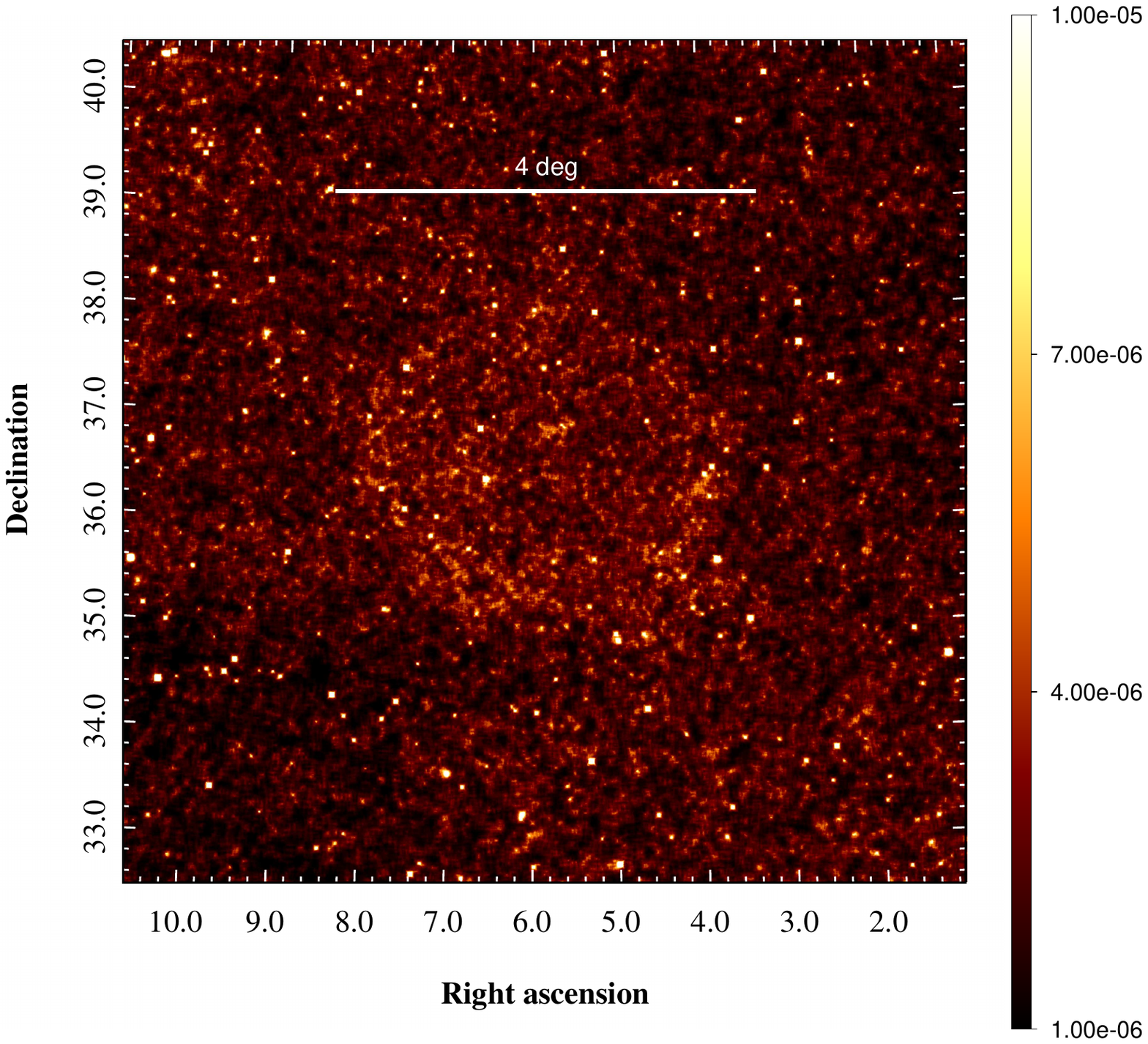}
\includegraphics[angle=0,bb=80 200 550 670,width=0.99\columnwidth]{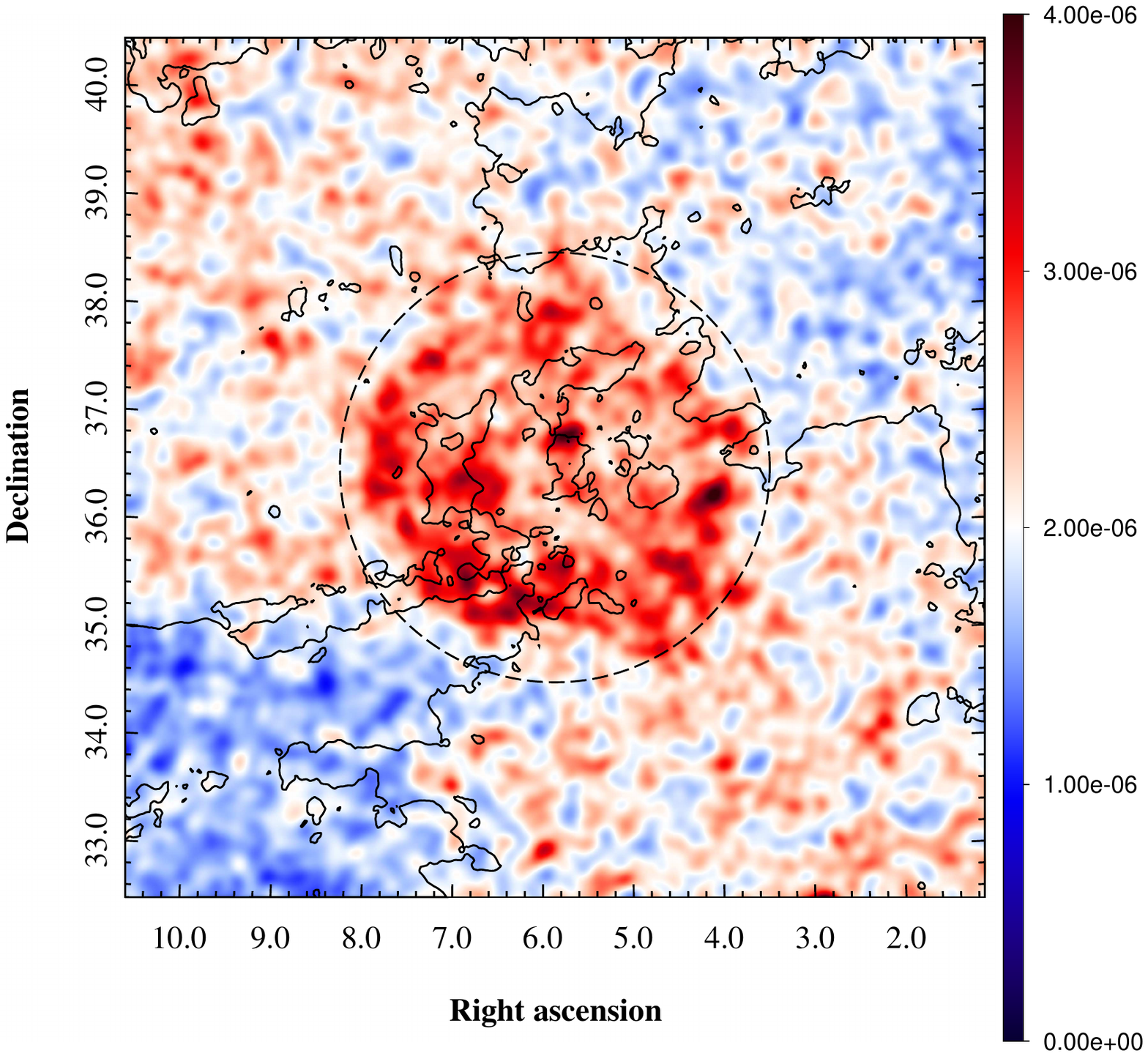} 
\caption{{\bf Left:} 0.5-0.7 keV image of the $8\times 8$ degrees region around  SRGe~J0023+3625 obtained during the first three all-sky surveys by \textit{SRG}/eROSITA. 
The surface brightness is in units of ${\rm counts\,s^{-1}}$ per $8"$ pixel per one (out of 7) eROSITA telescopes. {\bf Right:} The same image smoothed with a broad Gaussian ($\sigma=4'$) after excising compact sources with the X-ray flux above $10^{-13}\,{\rm erg\,s^{-1}\,cm^{-2}}$(0.5-2 keV).
Low surface brightness regions in the bottom-left and the top-right corners are due to the enhanced low energy absorption as illustrated by the contours of the dust emission 
}
\label{fig:image}
\end{figure*}

\begin{figure}
\centering
\includegraphics[angle=0,trim=1cm 5cm 0cm 4cm,width=0.95\columnwidth]{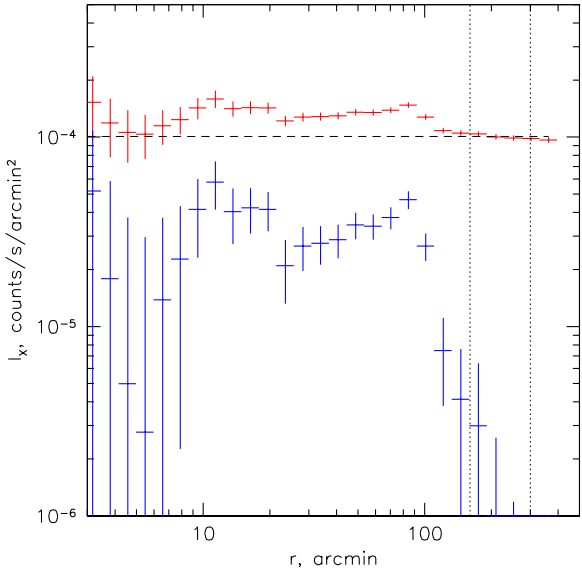}
\caption{Radial profile of the SNR candidate SRGe~J0023+3625 in the 0.5-0.7 keV band in units of ${\rm counts\,s^{-1}\,arcmin^2}$ per one eROSITA telescope unit. The top set of points (red) shows the total sky surface brightness. The bottom set of points (blue) show the same profile once the estimated sky background level (from the range of radii marked with dashed vertical lines is subtracted. The resulting profile is mildly edge brightened. There are (marginal) signs of brighter emission within the inner $\sim 20'$.  
}
\label{fig:radial}
\end{figure}

\begin{table}
\caption{Main observed parameters of the newly found diffuse object SRGe~J0023+3625.
}
\vspace{0.1cm}
\begin{center}
\begin{tabular}{rr}
\hline
 Parameter & Value  \\ 
\hline
\hline
Position (FK5, deg) & (5.79,36.42) \\
Position (Galactic, deg) & (116.6,-26.1) \\
Angular Size (Radius, $R_a$ deg)& $1.94$ \\
Total 0.3-2 keV Flux (${\rm erg\,s^{-1}\,cm^{-2}}$) & $\sim 3\times10^{-11}$ \\
\hline
\hline
\end{tabular}
\end{center}
\label{tab:obs}
\end{table}

\begin{figure}
\centering
\includegraphics[angle=0,trim=1cm 5cm 0cm 4cm,width=0.95\columnwidth]{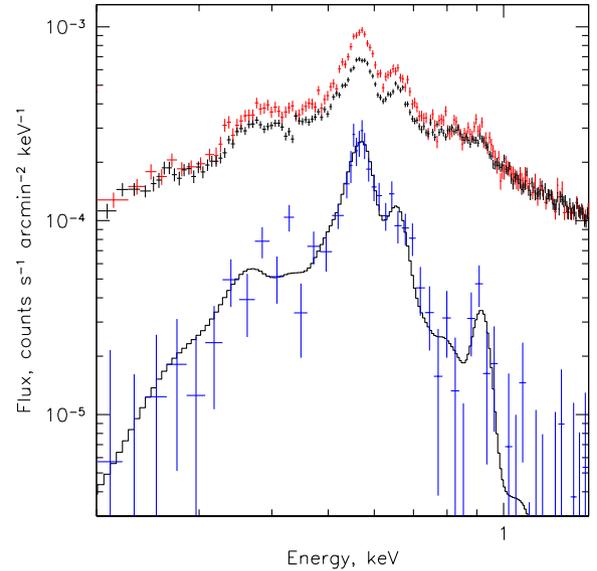}
\caption{Spectrum (in units of ${\rm counts\,s^{-1}\,keV^{-1}\,arcmin^{-2}}$) extracted from a {\bf circle} with radiius $1.95^\circ$ (red points). For comparison, the black points show the spectrum extracted from  an {\bf annulus} $r=1.95\div4^\circ$. The blue points show the difference between the red and black points, i.e., the excess emission associated with the newly found SNR candidate. The solid black curve shows the APEC CEI fit ($kT=0.17$~keV) to the SNR spectrum.
}
\label{fig:spec}
\end{figure}

The spectrum was extracted from a circle (radius $R_{a}=1.94^\circ$) after excising a few very bright compact sources with 0.5-2 keV flux above $10^{-13}\, {\rm erg\,s^{-1}}$. The corresponding spectrum corrected for the detector intrinsic background is shown in Fig.~\ref{fig:spec} with red points. For comparison, black points show the spectrum extracted from an annulus between $R_{a}$ and $4^\circ$. The blue points in this Figure show the difference between these two spectra, i.e. the excess emission inside the circle with the radius $R_{a}$. The spectrum of the SNR candidate is dominated by the oxygen lines similar to the surrounding sky spectrum, although the line ratios might be slightly different (see below). 
This spectrum can be reasonably well approximated by the absorbed \texttt{APEC} model with the temperature $0.17$~keV (see Table~\ref{tab:spec}). 

The absorbing column density of the model is consistent with the expected value of $N_H\sim 7.4\times10^{20}\,{\rm cm^{-2}}$ that is based on the HI column density in the HI4PI survey \citep[][]{2016A&A...594A.116H} and the amount of molecular gas traced by the dust in the extinction maps \citep[][]{2015ApJ...798...88M}.  The 3D reconstruction based on the Bayestar19 data \citep{2019ApJ...887...93G} show that bulk of the integrated absorption in that direction is accumulated over $\sim$300~pc from the Sun, suggesting that the newly found X-ray source is located farther away, although the value of the column density derived from the X-ray spectra might be model dependent. 

In the spectral model mentioned above, the abundance of all elements was set to the solar values, except for oxygen and neon that were free parameters of the model. For this temperature range, the emission is dominated by metals (rather than H and He bremsstrahlung) and, therefore, the model normalization and metals abundance are strongly degenerate. However, the best-fitting model suggests, though marginally, that neon is overabundant relative to oxygen. We discuss this issue later in \S\ref{sec:nei}. 

\begin{table}
\caption{Spectral parameters for the \texttt{tbabs*(v)apec} model, shown in Fig.~\ref{fig:spec} with the black line.}
\vspace{0.1cm}
\begin{center}
\begin{tabular}{rr}
\hline
 Parameter & Value  \\ 
\hline
\hline
$N_H\,{\rm [cm^{-2}]}$ & ($6.9\pm2.3\,)\times 10^{20}$ \\
$kT\,{\rm [keV]}$ & $0.17 \pm 0.009$ \\
Abundance, O & $0.8\pm 0.13$ \\
Abundance, Ne & $1.6\pm 0.6$ \\
$I_X,\, {\rm [phot\, s^{-1}\, cm^{-2}\, arcmin^{-2} ]}$ & ($1.1\,\pm 0.4\,) \times 10^{-6}$ \\ 
\hline
\hline
\end{tabular}
\end{center}
\label{tab:spec}
\end{table}

The observed X-ray properties of the newly found object can be summarized as follows: 
\begin{itemize}
\item It is a nearly circular, mildly edge-brightened object at $b\approx-26$.
\item Its surface brightness in the 0.5-0.7 keV band (dominated by OVII and OVIII lines) is a few tenths of the background sky brightness in the same energy band. 
\item The spectral shape is not dissimilar from the sky spectrum in the 0.3-1 keV  band (after removal of the CXB contribution), although the emission line helium-like neon seems to be more prominent in the SNR spectrum. 
\item Low energy absorption is consistent with the total column density of hydrogen atoms in that direction, which is presumably accumulated over 200-300 pc.
\end{itemize}

\subsection{Data at other wavelengths}

In order to find possible counterparts of this object at other wavelengths, we have checked publicly-available data on radio, optical, infrared and gamma-ray emission from that regions. We found no clear indication of the signal of similar extent and morphology in any of these maps, with probable exception of distorted ring-like structures visible in e.g. dust emission at 60 and 100 $\mu$m (based on the maps from the Improved Reprocessing of the IRAS Survey \citep[IRIS][]{2005ApJS..157..302M}. No correlating bright radio, optical or gamma-ray emission has been found. 

Next we discuss a few scenarios, which can explain some (or all) of these observational properties.


\section{Distant object scenario: SN~Ia remnant in the Galactic halo}
The measured column density of absorbing material that is consistent with the total column density of the Galaxy in this direction is best explained if the object is farther than $\sim$300~pc away from us. Coupled with the high galactic latitude it is plausible that the object is above the Galactic disk in the hot and low-density gas of the Milky Way halo. Below we argue that an SNI~a explosion in the Milky Way halo indeed provides us with an attractive explanation for the most salient features of the newly discovered object. 

\subsection{Basic idea}
Consider a SNIa explosion in a hot very low density environment ($n_{p,halo}\sim 10^{-3}\,{\rm cm^{-3}}$ and $T_{halo}\sim 2\,10^6\,{\rm K}$), characteristic for the hot and low density gas well above the Galactic Plane. This setup has a number of implications.

\begin{enumerate}
\item Due to the low density of the halo, the size of the SNR during the Sedov stage will be a factor a few larger than in the Galactic disk for a given SNR age \citep[an even more extreme version of such a setup, when an SN happens in a bubble of relativistic plasma inflated by an AGN in a galaxy cluster has been considered by][]{2011MNRAS.414..879C}. Compared to the ISM in the disk, the gas in the halo might have milder density variations on scales of 10-100~pc, so that the SNR could keep its approximately round shape for a longer time.
\item While the forward shock velocity in the halo will be larger (given the age or size of an SNR) than in the disk ISM, the Mach number of the forward shock can be low since the temperature upstream of the shock is relatively high. This might have an impact on the acceleration of particles at the forward shock and the generation of radio and high-energy emissions. Another possible implication is that adiabatic compression of protons and electrons might be the dominant process that sets the electron temperature downstream of the shock.
\item Due to extremely low density, the cooling time of the gas compressed by the shock is very long, so one can safely neglect the radiation losses of this gas. The time for reaching the collisional equilibrium is also long. This means that the gas downstream might 'remember' the halo CEI for a long time despite the increased temperature of the electrons. Such conditions should lead to a strong boost of the gas emissivity since hot electrons can easily excite transitions in ions, which are characteristic of a lower temperature plasma. As a result, such non-CEI gas might shine in the same lines as the halo but much more efficiently.
\end{enumerate}

We now consider a more quantitative example, which illustrates how these conditions i-iii might explain \textit{SRG} observations. The physical size of the newly discovered object is related to its distance $D$ as $R=R_aD=34 \left ( \frac{D}{\rm kpc}\right )\,{\rm pc}$. For the illustrative simulations, we assumed that $D=3\,{\rm kpc}$, so that $R\approx 100\,{\rm pc}$.

\subsection{Non-radiative numerical 1D model}

\begin{figure}
\centering
\includegraphics[angle=0,trim=1cm 5cm 0cm 4cm,width=0.95\columnwidth]{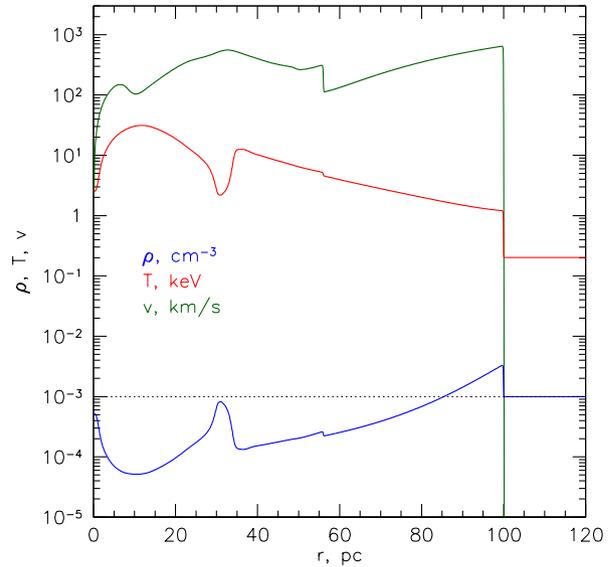}
\caption{Density, temperature and velocity profiles in non-radiative 1D simulations of a SNIa ejecta expansion in the low density gas of the Milky Way halo some 44000~yr after the explosion.}
\label{fig:model_profile}
\end{figure}

In order to illustrate the above model more quantitatively, we have performed a simple non-radiative 1D hydrodynamic simulation using publicly available hydro code \verb# PLUTO # \citep{2007ApJS..170..228M} under assumption of spherical symmetry. The homologously expanding envelope with the exponential density profile \citep[e.g.][]{1998ApJ...497..807D} is embedded into a homogeneous medium with the density $n_{p,halo}=10^{-3}\,{\rm cm^{-3}}$ and temperature $kT_{halo}=0.2\,{\rm keV}$. The total mass of the ejecta is $M=1.4\,M_\odot$ and the kinetic energy $E_K=1.3\,10^{51}\,{\rm erg}$. A snapshot of the model some $4.4\,10^4\,{\rm yr}$ after the explosion when the radius of the forward shock $R_{fs}\approx 100\,{\rm}$ pc, is shown in Fig.~\ref{fig:model_profile}.  
At this moment, the density compression factor at the shock is $C=\left ( \frac{\rho_d}{\rho_u}\right )\approx 3.3$, corresponding to the Mach number $M\approx 3.76$ in the gas with adiabatic index $\gamma=5/3$. The downstream temperature in the simulations, which do not distinguish electron and ion temperatures, is $\,\approx 1\,{\rm keV}$. We note here that a pure adiabatic compression would lead to a lower temperature $ C^{\gamma-1}\times T_{halo}\approx 0.44 \,{\rm keV}$. We, therefore, assume that the downstream electron temperature $T_{e,d}$ is somewhere between 0.44 and 1~keV. Below, we used  $kT_e=0.7\,{\rm keV}$ for estimates.

We note in passing that for display purposes the value of temperature shown in Fig.~\ref{fig:model_profile} is evaluated as the ratio of pressure $P$ and mass density $\rho_m$, which are the two quantities provided by the non-radiative single flluid hydrodynamic model.  Namely, $kT=\mu m_p P/\rho_m$,  where $\mu\approx 0.6$. This value of $\mu$ is appropriate for ionized plasma with Solar abundance of heavy elements. In our fiducial model, the mass of the gas swept by the forward shock is $\sim 100\,M_\odot$ when $R=100\,{\rm pc}$, i.e. much larger than the mass of the ejecta. Therefore, the above value of $\mu$ is suitable for outer regions, at least for estimates. This is also corroborated by approximately solar abundance of oxygen and neon in the SNR spectrum (see Tab.~\ref{tab:spec}).
However, for the ejecta, which are dominated by the heavy elements, the estimated temperature is likely far from either electron or ion temperatures.  The same applies to the gas number density $\rho$ shown in Fig.~\ref{fig:model_profile}, which is related to mass density $\rho_m$ as  $\rho=\rho_m/m_p$. 


A "bump" in the density distribution at $r\sim 30-35\,{\rm pc}$ is a contact discontinuity separating the SNR ejecta from the ISM swept by the forward shock. In our fiducial model, the reverse shock reaches the center some 15\,000~yr after the explosion. In terms of characteristic time $t_{ch}=M^{5/6}E_K^{-1/2}(n_p m_p)^{-1/3}\approx 5500\,{\rm yr}$, this moment corresponds to $\sim 3 t_{ch}$ in broad agreement with the analytic and numerical results of  \citet{1999ApJS..120..299T} that uses power law (or cored power law) density profiles. The 1D model predicts a reflection of the reverse shock at the center \citep[see, e.g.][]{1988ApJ...334..252C} and its transformation into a forward propagating (weak) shock. In Fig.\ref{fig:model_profile} this reflected shock is most easily seen in the velocity profile at $r\sim 55\,{\rm pc}$.

\subsection{Emission measure profile}

\begin{figure}
\centering
\includegraphics[angle=0,trim=1cm 5cm 0cm 4cm,width=0.95\columnwidth]{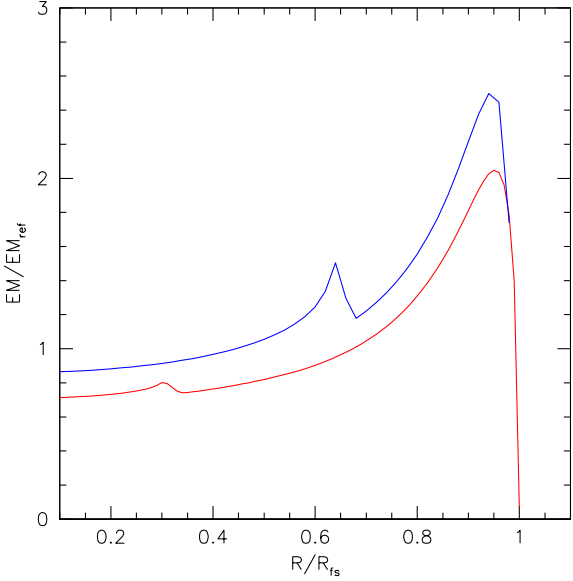}
\caption{Radial profile of the projected emission measure $EM=\int n_p^2 dl$ of an SNR, corresponding to the density profile shown in Fig.~\ref{fig:model_profile}. It is normalized by the forward shock radius (horizontal axis) and the reference value of the emission measure $EM_{ref}=2R_{fs}\times n_{p,halo}^2$ (vertical axis). The blue and red curves correspond to the moments when the forward shock was at 50 and 100~pc, respectively.}
\label{fig:em_sim}
\end{figure}

From the above simulations, one can generate a projected emission measure profile, namely, a line of sight (l.o.s.) integral of the density squared, $EM=\int n_p^2 dl$, shown in Fig.~\ref{fig:em_sim}. It is normalized by the forward shock radius (horizontal axis) and the reference value of the emission measure $EM_{ref}=2R_{fs}*n_{p,halo}^2$ (vertical axis), i.e. the l.o.s. emission measure of the halo gas through the diameter of the sphere with radius $R_{fs}$. Should the emissivity of the gas be constant, this emission measure profile could be directly compared with the properly normalized observed profiles shown in Fig.~\ref{fig:radial}. However, the constant emissivity is a  poor approximation. First of all, our simulations are non-radiative, which might be especially important for the inner part of the remnant swept by the reverse shock. Secondly, the abundance of heavy elements in the inner part might be dramatically different from halo abundance. Finally, equilibration time scales, in particular, the ionization equilibrium time scale, could be very long for the compressed halo gas.

For these parameters, the excess emission measure produced by the SNR would make only a few percent of the halo emission measure (see \S\ref{sec:discussion} below). We argue, however, that the emissivity of the shocked halo gas is strongly boosted due to the departure of the gas from CEI.

\subsection{Non Equilibrium Ionization}
\label{sec:nei}

The importance of Non Equilibrium Ionization for SNRs has long been recognized \citep[see, e.g.][]{1977PASJ...29..813I,1978A&A....65...99M,1978A&A....65..115M,1981ApJ...246L..27W,1982ApJ...262..308S}. For the context of our model, the problem greatly simplifies since the plasma is in a state when helium- and hydrogen-like ions\footnote{Ne-like ions for iron} of the most important elements dominate in the initial state (see below). In this case, the key features introduced by NEI can be easily understood. In fact, the closest examples of intensity/spectrum changes are discussed in \cite{1979PASJ...31...71H} and, in a concise form, are shown in fig.14 in the review of \citet{1999LNP...520..189L}. There, a spectrum of H+He+O plasma is plotted for CEI at $T=10^6\,{\rm K}$ together with the spectrum of the same gas, when electrons are instantaneously heated to  $T=6\,10^6\,{\rm K}$. From the comparison of these spectra, it is clear that the total X-ray flux increases by a large factor, while most of the prominent lines characteristic for $T=10^6\,{\rm K}$ plasma, remain in place.

As discussed above,  the electron temperature behind the forward shock is likely in the range of 0.44-1 keV. The lower limit corresponds to pure adiabatic compression, while the upper limit stands for fast equilibration of the electron and ion temperatures by some plasma process. For estimates, we will use some intermediate value $kT_e=0.7\,{\rm keV}$, although the arguments are valid for the entire temperature range of interest. For simplicity, we further ignore time variations of the electron temperature downstream of the shock.  

In our fiducial model, the halo gas is in CEI at $kT \approx 0.2\,{\rm keV}$. Accordingly, the fractions of OVII and OVIII ions are high (and comparable). When the temperature of electrons increases on the downstream side of the shock, the gas starts evolving towards CEI equilibrium at the new temperature (recall that we assume that electron temperature does not evolve after the instantaneous increase at the shock). If the new CEI (at $kT \approx 0.7\,{\rm keV}$) is reached, both OVII and OVII fractions will drop strongly (depending on the temperature), with the most dramatic changes for the He-like oxygen ions. The new equilibrium state will be achieved after time $t_{i}\approx [n_e\sigma_i(T_e)]^{-1}$, where for our fiducial run, $n_e\approx n_p\approx 3\,10^{-3}\,{cm^{-3}}$, and $\sigma_i(T_e)$ is the ionization cross section of either OVII or OVIII. 

For this study, it is sufficient to calculate $t_{i}$ for OVII ions, since their ionization cross section is larger. For the temperature range of interest $\sigma_{i,ovii}$ changes between $2\,10^{-11}$ to $8\,10^{-11}\,{\rm cm^{-2}}$ \citep[e.g.][]{2007A&A...466..771D}. Therefore, the ionization balance will change strongly on time scales longer than $t_{i}\approx 1.3\,10^5\,{\rm yr}$, that is longer than the age of the SNR in our fiducial run. Therefore, one can expect that the ion fractions will stay unchanged, despite high electron temperature, while the excitation rate will be boosted by the elevated temperature, primarily because the factor $e^{-\Delta E/kT_e}$ in the cross section is now close to unity. As the result, one can expect to find the same set of the lines as in the halo spectrum but brighter emission from the downstream region.

The above qualitative discussion can be illustrated more quantitatively, using existing NEI models. It turns out, that for our case (a sudden increase of temperature from the initially hot state) the most suitable is the \texttt{rnei} model \citep{2001ApJ...548..820B} available in \texttt{XSPEC}. While originally intended for recombining/cooling plasma, it appears to work well for our case too. In our context, this model has three important parameters: initial plasma temperature $T_b$, which sets the initial CEI, a new plasma temperature $T_f$ (corresponds to downstream electron temperature), and the ionization time scale $\tau=t\times n_e$, where $t$ is the time elapsed after the temperature change. In our fiducial model here, $kT_b=0.15\,{\rm keV}$,  $kT_f=0.5\,{\rm keV}$,  $n_e\sim 3\,10^{-3}\,{\rm cm^{-3}}$, $t=4.4\,10^{4}\,{\rm yr}$ and $\tau\sim 4\,10^9\,{\rm s\,cm^{-3}}$. 

\begin{figure}
\centering
\includegraphics[angle=0,bb=30 180 580 670, width=0.95\columnwidth]{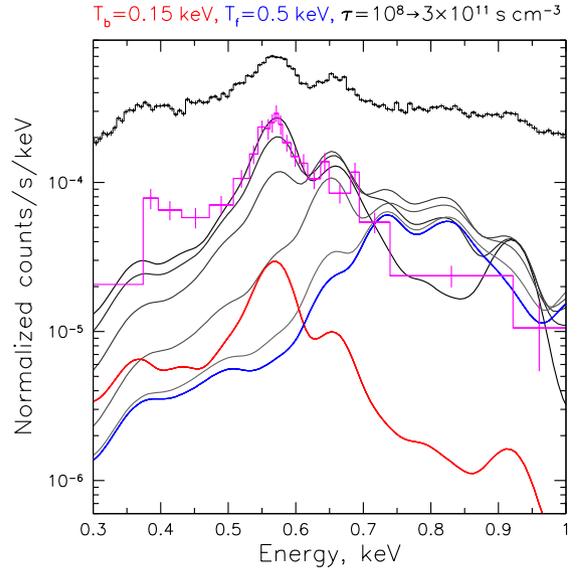}
\caption{Spectra predicted by the \texttt{rnei} model (gray curves), when the temperature changes abruptly from $kT_i=0.15\,{\rm keV}$ to $kT_n=0.5\,{\rm keV}$. The spectra have been convolved with the eROSITA response. Different gray curves correspond to different values of the parameter $\tau=t\times n_e$: $10^8$, $10^{10}$, $3\times10^{10}$,$10^{11}$, and $3\times10^{11}$ ${\rm s\,cm^{-3}}$ (from darker to lighter gray). 
The "red" and "blue" spectra show the CEI emission for $kT=kT_b$ and $kT_f$, respectively. The black histogram shows the measured spectrum extracted from the background region, while the magenta histogram with data points show the spectrum extracted from the SNR region after subtracting the contribution of the background emission. This plot shows (i) almost an order of magnitude increase of the emissivity due to the temperature increase (compare top gray curve with the "red" spectrum) and (ii) resemblance of the NEI spectrum shape at small $\tau$ to the CEI spectrum at the lower initial temperature. These are the two key signatures needed to explain observations. 
}
\label{fig:rnei_ti015tf05tau9to115}
\end{figure}

\begin{figure}
\centering
\includegraphics[angle=0,trim=0cm 5cm 0cm 3cm,width=0.95\columnwidth]{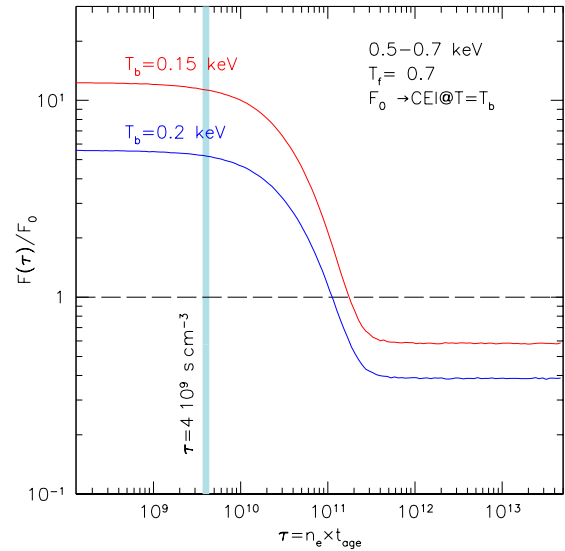}
\caption{0.5-0.7~keV emissivity as a function of $\tau=n_e\times t$ for two values of the gas initial temperature $T_b=0.15$ and 0.2~keV, respectively. The temperature of electrons is 0.7 keV in both cases. The emissivity is normalized by the corresponding value for the CEI case at $T=T_b$. The vertical blue line is the fiducial value of $\tau$ in our simulations. 
}
\label{fig:rnei_flux_tau}
\end{figure}

A set of spectra predicted by the \texttt{rnei} model is shown in Fig.~\ref{fig:rnei_ti015tf05tau9to115}. The "red" and "blue" spectra show the CEI emission for $kT=kT_b$ and $kT_f$, respectively. As expected, the change of the temperature from 0.15 to 0.5 keV dramatically changes the spectrum emitted by plasma in CEI. In particular, the OVII and OVIII disappear from the spectrum and the majority of photons are now coming in the 0.7-1 keV band. In contrast, the top gray line shows the spectrum emitted by the same plasma soon after an abrupt change in the temperature, before the ionization equilibrium adjusts to the new temperature. In terms of the  \texttt{rnei} parameters, the top gray curve corresponds to $\tau=10^8 \,{\rm s\,cm^{-3}}$. Other (lower) gray curves correspond to the monotonically increasing value of $\tau$ that changes from $10^{10}$ to $3\,10^{11}\,{\rm s\,cm^{-3}}$. As expected, the emissivity is boosted by a large factor (almost an order of magnitude!) by the sudden temperature increase. When $\tau$ is less than $\sim 10^{10}\,{\rm s\,cm^{-3}}$, the emissivity remains strongly boosted, but the contribution of the O~VII line slowly decreases. At larger $\tau$ (lowest three gray curves correspond to $\tau=3\,10^{10},\,10^{11},\,3\,10^{11}$), O~VII disappears, emissivity drops and the spectrum approaches CEI spectrum for $kT_f$. The increase of emissivity is further illustrated in Fig.~\ref{fig:rnei_flux_tau} that shows the 0.5-0.7 keV emissivity as a function of $\tau$ for two values of the initial temperature $T_b$. The emissivity is normalized to its value for CEI at $T=T_b$. 

From Figs.~\ref{fig:rnei_ti015tf05tau9to115} and  \ref{fig:rnei_flux_tau} it is clear that (i) NEI effects indeed boost the emissivity in the oxygen lines by an order of magnitude and (ii) the emitted spectrum from the $kT\sim 0.4-0.7\,{\rm keV}$ plasma resembles CEI spectrum with $kT\sim 0.15-0.2\,{\rm keV}$. This conclusion is valid for $\tau$ less than $\sim 10^{10}\,{\rm s\,cm^{-3}}$. For comparison, our fiducial simulations correspond to $\tau\sim 4\,10^{9}\,{\rm s\,cm^{-3}}$. A more detailed comparison shown in Fig.~\ref{fig:rnei_ti015tf05tau9to115} corroborates this conclusion. 

While the downstream spectrum features lines characteristic for lower temperature gas (due to NEI), the line ratios differ from the CEI cases either at $T=T_b$ or $T=T_f$. This allows one to measure both temperatures if the line ratios can be robustly measured. This is illustrated in  Fig.~\ref{fig:rnei_tbtf} that compares predicted and observed flux ratios. To make the comparison more robust, we have calculated fluxes in several moderately narrow bands, rather than fitting line fluxes with narrow Gaussians. These energy bands are given in Table~\ref{tab:bands}. Red, blue and green colors in Fig.~\ref{fig:rnei_tbtf} correspond to O~VIII, Ne~IX and Fe~XVII bands, respectively (see Tab.~\ref{tab:bands}). The fluxes in individual bands are normalized by the flux in the "O~VII band". The horizontal lines show observed values. The dashed curves show predicted ratios for the same bands in the \texttt{rnei} model where the electron temperature $T_f$ is fixed at 0.7~keV, while the pre-shock temperature varies between 0.1 and 0.7 keV. Circles show intersections of these curves with the observed ratios, suggesting $T_b\sim 0.15$~keV. Qualitatively, this value of temperature is driven by the weakness of O~VIII and Fe~XVII lines in the observed spectrum. 

Similarly, the solid curves show expected line ratios when $T_b$ is fixed at 0.15~keV, while $T_f$  varies between 0.1 and 0.7 keV. The intersections of these curves with the observed ratios suggest $T_f\sim 0.4$~keV, although the accuracy of this estimate is limited since the line ratio mildly depends on $T_f$ when the transition energy is close to the O~VII line. The Ne~IX line has the largest energy and, therefore, the corresponding curve is more affected. Therefore, when fitting such spectrum with the CEI APEC model, one can expect to find an overabundance of Ne relative to O, as indeed observed from this SNR (see \S\ref{sec:obs}). This, however, is not a watertight argument, since it is possible that in the standard abundance table used here \citep{1989GeCoA..53..197A}, the abundance of neon relative to oxygen is underestimated by a factor of $\sim$2.5 \citep{2005Natur.436..525D}. If true, this would weaken the evidence of the NEI effects in the observed spectrum.

\begin{table}
\caption{Energy bands used for crude diagnostic of the upstream ($T_b$) and downstream ($T_f$) temperatures.
}
\vspace{0.1cm}
\begin{center}
\begin{tabular}{rrr}
\hline
Band ID & Energy (keV) & Observed Flux  \\ 
& & ${\rm counts\,s^{-1}\,arcmin^{-2}}$ per tel.\\
\hline
\hline
O~VII & 0.52--0.61 & 0.58\\
O~VIII & 0.61--0.70& 0.32\\
Ne~IX & 0.87--0.96& 0.092\\
Fe~XVII & 0.78--0.87& 0.044\\
\hline
\hline
\end{tabular}
\end{center}
\label{tab:bands}
\end{table}


\begin{figure}
\centering
\includegraphics[angle=0,trim=0cm 5cm 0cm 4cm,width=0.95\columnwidth]{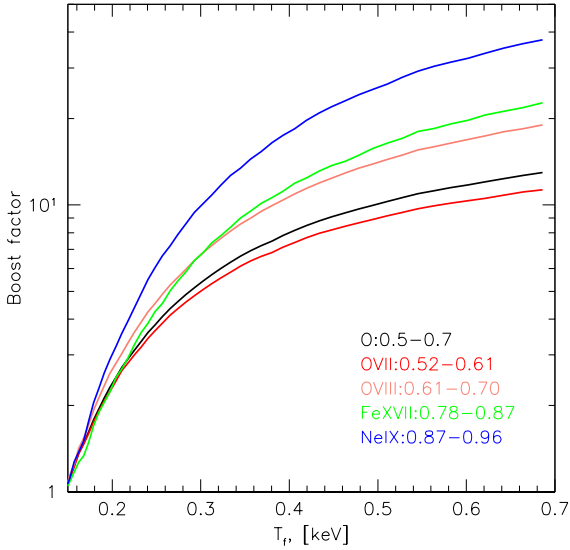}
\caption{Expected increase of the flux in a given energy band (listed in Table \ref{tab:bands}) when the electron temperature  changes instantaneously from $T_b=0.15\,{\rm keV}$ to $T_f$ based on the \texttt{rnei} model. The boost factor is calculated relative to the flux in the CEI model at $T=T_b$.   
}
\label{fig:rnei_boost}
\end{figure}

\begin{figure}
\centering
\includegraphics[angle=0,trim=0cm 5cm 0cm 4cm,width=0.95\columnwidth]{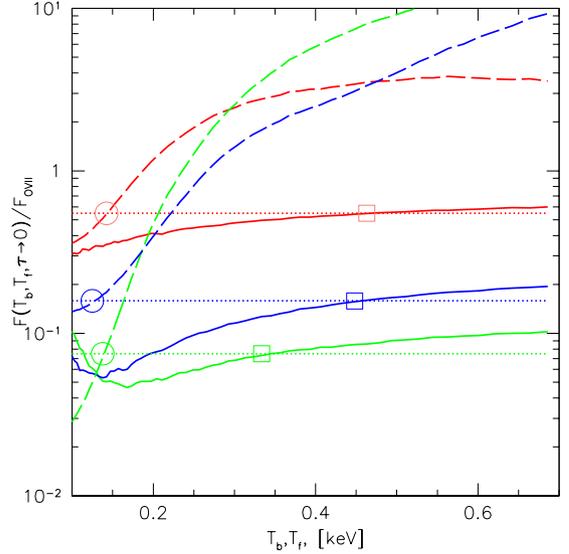}
\caption{Constraints on the preshock and post-shock temperatures derived from the line ratios under the assumption of  $\tau\ll 10^{10}$. Red, blue and green colors correspond to O~VIII, Ne~IX, and Fe~XVII bands, respectively (see Tab.~\ref{tab:bands}). Fluxes are normalized by the flux in the "O~VII band". The horizontal lines show observed values. The dashed curves show predicted ratios for the same bands in the \texttt{rnei} model where the electron temperature $T_f$ is fixed at 0.7~keV, while the pre-shock temperature varies between 0.1 and 0.7 keV. Circles show intersections of these curves with the observed ratios, suggesting $T_f\sim 0.15$~keV. Similarly, the solid curves show expected line ratios when $T_b$ is fixed at 0.15~keV, while $T_f$  varies between 0.1 and 0.7 keV. The intersections of these curves with the observed ratios suggest $T_f\sim 0.4$~keV, although the accuracy of this estimate is limited.
}
\label{fig:rnei_tbtf}
\end{figure}


\subsection{Constraints from radio and GeV-TeV bands} 
\begin{figure}
\centering
\includegraphics[angle=0,bb=40 200 530 620,width=0.99\columnwidth]{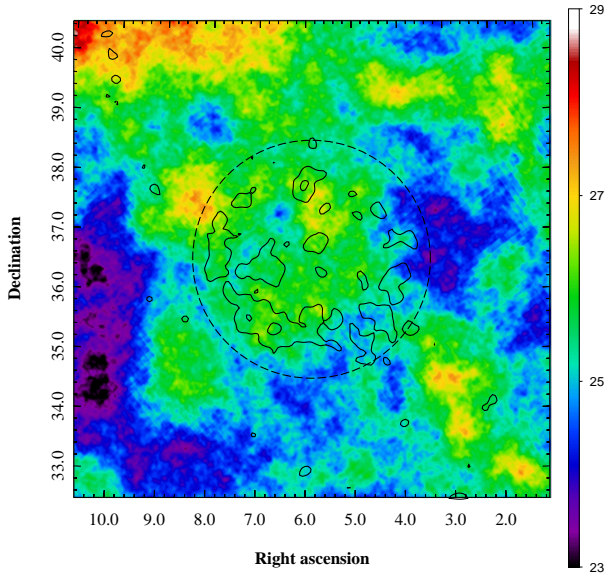}
\caption{The 408 MHz map of the source vicinity taken from the all-sky radio survey of \citet{Haslam82} reprocessed and re-evaluated by \citet{Remazeilles_408MHz}. The black contours corresponds to X-ray observations of SRG discussed above.}  
\label{fig:haslam}
\end{figure}

The nearby Type Ia supernovae like Tycho's SNR, SN 1006 are known to be sources of non-thermal radiation from radio to gamma-rays \citep[see e.g.][]{vink20}. Consider SN 1006 as the generic example of the high altitude Type Ia supernova. SN 1006 has the apparent size of about 30 arcminutes with the radius $R$ of about 9.5~pc at the estimated distance of 2.2~kpc. The SNR is located  14 degrees (likely height 550 pc) above the galactic plane  and from the deep X-ray observations of the XMM-Newton  the ambient density in the southeastern limb was estimated to be 0.035 cm$^{-3}$ with the shock velocity of about 5000 km s$^{-1}$
\citep[see e.g.][and the references therein]{Miceli16}. The bilateral structure which is apparent in radio and X-rays can be explained by the high efficiency of particle acceleration in the bright sections. Non-thermal radio and X-ray emission from NE and SW limbs revealed synchrotron radiation produced by TeV regime electrons accelerated at the supernova forward shock \citep[see e.g.][]{Helder12}. 

The gamma-ray emission from SN 1006 was detected by High Energy Stereoscopic System (H.E.S.S.) above 1 TeV at the flux level $> 2\times 10^{-13}$ cm$^2$s$^{-1}$ and the power-law photon index close to 2.3 \citep[][]{HESS10}. Fermi LAT detected SN 1006 at 6 $\sigma$ significance level  
\citep[][]{Fermi_Xing16,Fermi_Condon17} with a photon spectral index of $\sim$ 1.8. The indications of asymmetry between the NE and SW limbs were found. The northeast limb was firmly detected while no significant detection of the southwest limb was established. The energy flux about 5$\times 10^{-12}$ erg cm$^2$ s$^{-1}$ in the energy range between 1 GeV and 2 TeV was derived. Modeling of the gamma-ray emission of SN 1006 suggested both leptonic and hadronic contributions into the observed flux \citep[e.g.][]{Miceli16,Fermi_Condon17,Winner20}.To explain the observed profiles of the synchrotron X-ray emission  \citet{Miceli16} suggested an interaction of the SW part of SN 1006 with an atomic cloud of a number density $\sim$ 0.5 cm$^{-3}$. The best fit models developed by \citet{Winner20} suggested the uniform ambient number density of 0.12 cm$^{-3}$ which they find to be compatible with the existing multi-wavelength observations. All of the number density estimations  
suggested that even an SNR of the same age and energy as SN 1006 being located at the distance 3 kpc in the low density plasma $\sim$ 10$^{-3}$ cm$^{-3}$ can not be detected by Fermi LAT or by the current generation of the imaging atmospheric  Cherenkov telescopes. An SNR of some 40,000 years old age would not likely accelerate TeV regime particles by diffusive shock acceleration. The high energy nuclei accelerated at the earlier stages at fast forward shock would escape from the source while the lower energy cosmic rays confined inside the SNR would suffer from strong adiabatic deceleration when SNR expanded to the radius of $\sim$ 100 pc.

GeV regime electrons can still be accelerated by the SNR forward shock of a speed about 500 km s$^{-1}$. The diffusive shock acceleration model predicts the index of the power-law distribution of particle momenta dN $\propto p^{- \alpha}$dp to be $\alpha = (C+2)/(C-1)$ for the shock compression ratio $C$ \citep[e.g.][]{BE87}. The index $\alpha \approx 2.3$ for $C=3.3$ and the corresponding index of the radio synchrotron flux is about 0.65.   

To estimate the expected radio flux of Type Ia SNR one may use the well known relation between the surface brightness $\Sigma_{\nu}$ at a frequency $\nu$ and the SNR diameter $D$  which can be expressed as $\Sigma_{\nu} = A \cdot D^{-\beta}$ \citep[see e.g.][]{L81,B86,BV04}. Where the factor $A$ depends on the energy and mass
of the SN ejecta, the density of the ambient matter and magnetic fields, while the index $\beta$ is only weakly dependent on the parameters \citep[see e.g.][and the references therein]{sigma_D05}. While the use of $\Sigma_{\nu}-D$  relation to derive the SNR diameters and distances from the measured $\Sigma_{\nu}$ values may be rather uncertain \citep[see e.g.][]{Green04} it can still be used with some care to estimate the expected evolution of the radio brightness with SNR diameter based on the SNR synchrotron emission modeling  \citep[see e.g.][]{BV04}. 

The observed surface brightness $\Sigma_{\nu} \approx 3.2 \times 10^{-21}$ W m$^{-2}$ Hz$^{-1}$ sr$^{-1}$ \citep[see][]{sigma_D05} at $\nu$ = 1 GHz.  Applying the  $\Sigma_{\nu}-D$ relation to the generic case of SN 1006 one can estimate its surface brightness for radius of 100 pc to be below 
$3 \times 10^{-24}$ W m$^{-2}$ Hz$^{-1}$ sr$^{-1}$ assuming a conservative index  $\beta \geq 3$. Assuming the linear dependence of the factor $A$ on the ambient density the lower surface brightness  below $3 \times 10^{-26}$ W m$^{-2}$ Hz$^{-1}$ sr$^{-1}$ can be expected.

The estimations above which were made for SN 1006-like object can explain the apparent lack of the radio counterpart in the maps of the 408 MHz survey by \citet{Haslam82} while the conservative upper limit is not too much lower than the fluxes shown in Fig.~\ref{fig:haslam}\footnote{\citet{Remazeilles_408MHz} re-evaluated and re-processed the rawest 408 MHz data by \citet{Haslam82} to produce an improved source-subtracted and de-striped 408 MHz all-sky map which we used to make the map.}. Despite the large uncertainties in the index $\beta$ and in the dependence of $A$ on the ambient density, which may decrease the estimate of the radio flux,  new sensitive low frequency observations with {\sl LOFAR} \citep{LOFAR} might help to clarify the situation, although the source position is not the most favorable for this facility. The source is a good target for the future {\sl Square Kilometer Array}.

\subsection{Expected number of Type Ia SNRs at $z > 1$ kpc}

At heights over Galactic plane $|z| \geq 1$ kpc the stellar population is composed 
  primarily of the halo and thick disk \citep{Bland-Hawthorn_2016}. 
The age of the stellar halo is $\sim10$ Gyr \citep{Deason_2019};  most stars of the thick disk are probably of the same age \citep{Helmi_2020}. 
Both populations are therefore as old as the stars of elliptical galaxies, which
  prompts us a possibility to use the specific rate of SNe~Ia in E-galaxies 
  to estimate the expected rate of SN~Ia in the halo and the thick disk.
  
The observational estimate of the SN\,Ia rate for E-galaxies with stellar mass $ < 10^{11}$\msun\ is $\sim 0.09$ SN\,Ia per 100\,yr per $10^{10}$\msun\ (Li et al. 2011). 
Adopting the same specific rate for the MW halo with the stellar mass of 
  $1.4\times10^9$\msun\ \citep{Deason_2019} and the thick disk with the mass of $6\times10^9$\msun\, we get the SN\,Ia rate of $\sim 1.3\times10^{-4}$ SN\,Ia\,yr$^{-1}$ in the halo and of $\sim 5.4\times10^{-4}$ SN\,Ia\,yr$^{-1}$ in the thick disk. The current number of old SN\,Ia SNR with the age of $\leq 10^5$ yr is then of $\sim 13$ in the halo and $\sim 54$ in the thick disk. 
  
  We are primarily interested in SNRs in the low-density hot gas, i.e. well above the Galactic Plane, say $z\gtrsim$1~kpc, where $z$ is the distance above the Plane. 
 To get insight into the SNR number at $z \geq 1$ kpc we use Monte Carlo technique based on the stellar density distribution in halo and thick disk. For the halo following \citep{Deason_2019}  we adopt spheroidal distribution along the 
   galactocentric distance $r^2 = R^2 + (z/q)^2$ ($q = b/a = 0.6$) and a broken power law profile 
\begin{equation}
\rho \propto r^{-\beta}\,,
\end{equation}
  where $\beta = 2.3$ for $R \leq 27$ kpc and $\beta = 4.6$ for $R \geq 27$ kpc. 
For the thick disk the stellar density distribution is taken in the form 
  of multiplicative exponential law $f \propto \exp{(-R/h_R)}\exp{(-z/h_z)}$ with 
    $h_z = 0.9$ kpc and $h_R = 2.1$ kpc \citep{Bland-Hawthorn_2016}.



\begin{figure}
\centering
\includegraphics[angle=0,trim=0cm 5cm 0cm 11cm,width=0.99\columnwidth]{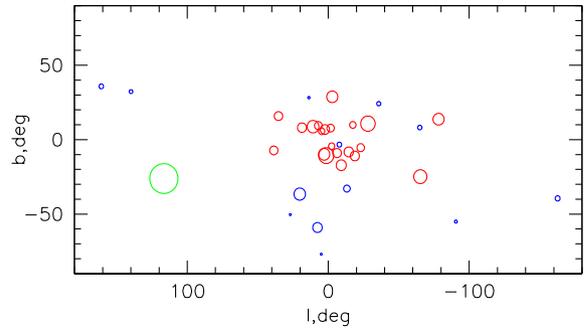}
\caption{Random Monte-Carlo realization of old and off-plane Type Ia SNRs (age $\leq 10^5$~yr; $z>1\,{\rm kpc}$) distribution over the sky, assuming that SNRs trace
  stellar density in the halo ({\em blue}) and thick disk ({\em red}). The size of each circle is inversely proportional to the distance of the SNR from the Earth, but does not correspond to the "observed" angular size. The green circle shows G116.6-26.1. Due to stochastic nature of the plot, one can only conclude that (i) there should be some 20-30 such objects in the Galaxy and (ii) the objects as close as G116.6-26.1 are rare.
}
\label{fig:snr_mcmap}
\end{figure}

The (random) positions of halo and thick disk SNRs with $z > 1$ in a Monte Carlo run 
  is shown in Figure \ref{fig:snr_mcmap}. Within the heliocentric sphere of the radius 
   $d\sim 4$~kpc one expects to find one old SNR at the heights $z > 1$ kpc. This demonstrates that the objects with parameters of G116.6-26.1 ($d \approx 3$ kpc and $z \sim 1.3$)  are very rare, but, nevertheless, not to the extent that would make the association with G116.6-26.1 implausible.

 
\section{Nearby object}

Although the inferred absorbing column density $N_H \sim 7\times10^{20}$ cm$^{-2}$ does not favor the nearby ($d < 300$ pc) SNR scenario, we find it sensible to mention two interesting alternatives to the origin of the found object: a nova shell and a shell related to a failed supernova.
  
\subsection{Nearby nova shell}

Adopting the nova shell initial velocity of 3200\kms\ comparable with that of 
  the fast nova V1721 Aql \citep{Hounsell_2011} and a typical ejecta mass  
  of $10^{-5}$\msun\,  one expects the shell kinetic energy of $E \approx 10^{45}$\,erg.
This value is predicted for the nova ejecta in the case of the thermonuclear runaway 
  in the accreted shell on the surface of 1\msun white dwarf \citep{Epelstain_2007}.
The present-day expansion velocity of $500$\kms\ implies the swept-up mass 
  $M_{sw} = 2E/v^2 = 4\times10^{-4}$\msun, and a moderate radius  of $0.16n^{-1/3}$\,pc,
  where $n$ is the ambient density.
This estimate combined with the angular radius of $2^{\circ}$ implies the small distance, well inside the Local Bubble (LB) of the $\sim 100$\,pc radius \citep{2009Ap&SS.323....1W}.
For the LB density $n = 0.01$\cmq\ \citep{Farhang_2019} one gets $r = 0.7$\,pc 
 and the distance $d = 20$\,pc.
The Sedov expansion law  $r = (E/\rho)^{1/5}t^{2/5}$ suggests the shell age $t = 890$\,yr.

The nova scenario implies the presence of the binary with the accreting white dwarf close to the shell center. According to the database of International Variable Star Index ({\em www.aavso/vsx}) no cataclysmic variable is present inside the circle with the radius of $1^{\circ}$, which casts doubt on this scenario. 

\subsection{Remnant of a nearby failed supernova}

\citet{Nadezhin_1980} recognized that the core collapse into a black hole that 
  avoids explosion nevertheless results in the ejection of low energy 
  envelope caused by the gravity weakening due to the neutrino emission.
If a pre-supernova is a red super-giant this mechanism can eject several solar mass 
   with the velocity of about 100\kms. 
In the case of a compact pre-supernova
  (blue super-giant or WR star) the Nadyozhin mechanism results in the low-mass 
  high-velocity ejecta \citep{2021PASJ...73L...6T}.   
The WR star as a presupernova is of particular interest since in this case the model 
  predicts ejecta with very low mass $5\times10^{-4}$\msun, high velocity,  
  $2000$\kms, and the kinetic energy of $3\times10^{46}$\,erg  \citep{2021PASJ...73L...6T}.
For the ambient density $n = 0.01$\cmq\ and the present-day expansion velocity of 500\kms\ the swept-up mass is 0.012\msun,  the shell radius is $r = 2.2$\,pc, 
  the age is $t = 2700$\,yr, and the distance is $d = 62$\,pc, i.e. the shell
  indeed is inside the LB.
   
The scenario of failed supernova suggests that at the center of the X-ray shell 
 there should be a black hole with the mass of the fully collapsed WR star, i.e., 
 $M_{bh} \sim 10$\msun.  One can imagine at least two scenarios, when the resulting black hole remains a source of X-rays long after the supernova explosion.
In one scenario, the black hole is powered by fallback accretion \citep[e.g.][]{2018MNRAS.476.2366F,2020ApJ...897L..44T}, while in the other - it is the Bondi accretion of the hot gas. In the former scenario, the luminosity of the source depends critically on the extrapolation of the accretion rate decline rate over thousands of years.  
In the latter case, a crude estimate of the expected accretion luminosity can be found assuming a fiducial value for the hot gas sound speed of 500\kms\ and the density 
  $n = 0.01$\cmq\, in which case  
  $\dot{M} \approx  2.9\times10^6(M_{bh}/10_{\odot})^2$\,g\,s$^{-1}$.
The corresponding luminosity is  $L_x = \eta\dot{M}c^2 \approx 2.6\times10^{27}\eta$\ergs, 
  where the radiation efficiency $\eta \lesssim 10^{-6}$ given $\dot{M}\ll L_E/c^2$ 
  \citep{Park_2017}.
For $\eta = 10^{-6}$ one gets $L_x \sim 2\times10^{21}$\ergs\ and the flux  
  of $\sim 10^{-20}$\,erg\,cm$^{-2}$\,s$^{-1}$, which is beyond the detection
  capabilities of the current generation of X-ray telescopes. It is clear that both scenarios have very large uncertainties and the lack of X-ray bright object in the center of the SNR does not immediately exclude the presence of a massive black hole there.
 However, at such small distances, parallax measurements might be within reach in the optical or radio bands, provided that bright and sharp features are identified in the SNR.

  \subsection{'Regular' nearby SNR}
  
\begin{figure}
\centering
\includegraphics[angle=0,bb=90 180 530 620,width=0.9\columnwidth]{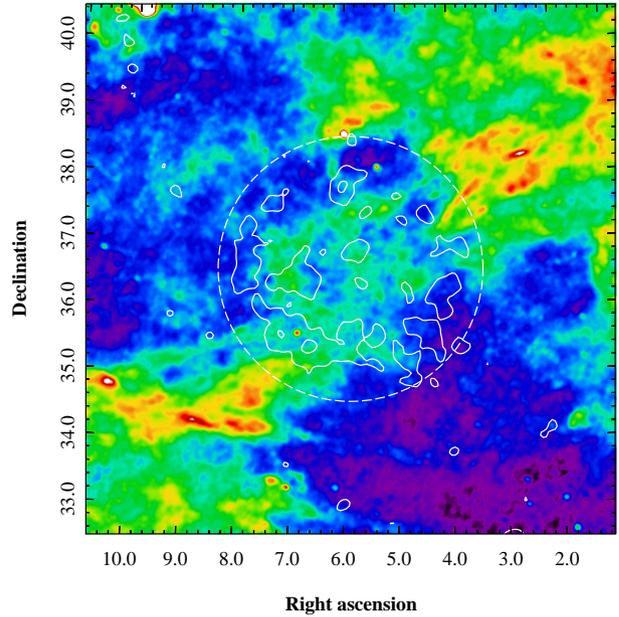}
\caption{IRAS 100~$\mu$m map with the X-ray contours superposed. Both X-ray and IR images possess complicated substructures, which show signs of correlation/anti-correlation. A certain degree of anti-correlation is expected due to photoelectric absorption of a distant X-ray source. However, the appearance of the dust map leaves open a possibility that the dust distribution is "affected" by  G116.6-26.1. If true, this could become an argument in favor of a local (less than 300~pc) SNR.}
\label{fig:iras}
\end{figure}

  Finally, there is a possibility that G116.6-26.1 is a "regular" type II SNR located a few 100 pc from us. One can find some evidence (although not compelling) for this scenario when considering signs of correlations between the structures seen in X-ray and IR images. We illustrate this point in Fig.~\ref{fig:iras} that shows the IRAS image together with the contours of X-ray emission associated with  G116.6-26.1.
  
  Some degree of anti-correlation between these two images is expected for any extended X-ray object that is farther away than the dust. However, the appearance of the dust map leaves open a possibility that the dust distribution is "affected" by  G116.6-26.1. If true, this would imply that the newly found object is co-spatial with the dust, i.e. within $300$~pc from us, as follows from the extinction distribution in the 3D Bayestar-2019 maps \citep{2019ApJ...887...93G}. In this case, some of the absorbing gas might be in front and some behind the extended X-ray source.  It should be  noted here that known nearby SNRs with radius $\sim$ 10 pc typically demonstrate detectable non-thermal radio and gamma-ray emission \citep[see e.g.][]{Helder12}, which is not yet identified in G116.6-26.1.
  
  We leave the discussion of this scenario for future studies and in the remaining part of the paper focus on the distant SNIa model.


\section{Discussion}
\label{sec:discussion}



An explosion of a supernova in a hot and very low density medium represents an interesting case when radiative cooling of the compressed ISM can be safely ignored. A supernova, exploding in a "normal" ISM  (with the density $\sim 1\,{\rm cm^{-3}}$) goes through three main phases: (i) free expansion, (ii) Sedov-Taylor phase, and (iii) momentum-conserving Snow-plow phase initiated by the rapid cooling that forms a thin dense shell \citep[see, e.g.][for recent simulations]{2021MNRAS.504..583S}. For the very low density medium, the SNR will dissolve in the ambient medium before entering the latter phase \citep[e.g.][]{1988ApJ...334..252C}.
 Indeed, $t_{cool}=\frac{3}{2}\frac{kT}{n\Lambda(T)}$ is longer than $10^8\,{\rm yr}$ for $n\sim 3\,10^{-3}\,{\rm cm^{-3}}$ for the relevant temperatures and Solar abundance of heavy elements. Here, $\Lambda(T)$ is the cooling function of the gas in CEI. Accounting for enhanced radiative losses due to NEI effects does not change this estimate, since CEI is restored on shorter time scales. As a result, the dense shell never forms and the SNR continues to stay on the non-radiative Sedov-Taylor type solution\footnote{We note in passing that in our  model the effects of magnetic fields and cosmic rays are completely neglected, although they might affect the dynamics of the SNR expansion at some stages of its evolution.} until the expansion velocity approaches the sound speed of the external medium, the compression ratio goes down and the SNR becomes indistinguishable from the ambient gas \citep[e.g.][]{1988ApJ...334..252C}. This condition sets the maximal {\it observable} size of the SNR - $R_{max}\sim 151 \left ( \frac{E_{51}}{P_4 \beta^2}\right )^{1/3}\,{\rm pc}$ \citep[eq. 4.9 in][]{1988ApJ...334..252C}, where $E_{51}$ is the energy of the explosion in units of $10^{51}\,{\rm erg}$, $P_4=\frac{nkT}{10^{4}\,{\rm K cm^{-3}}}$, $\beta$ is a parameter of order unity. 



\begin{figure}
\centering
\includegraphics[angle=0,trim=1cm 5cm 0cm 4cm,width=0.95\columnwidth]{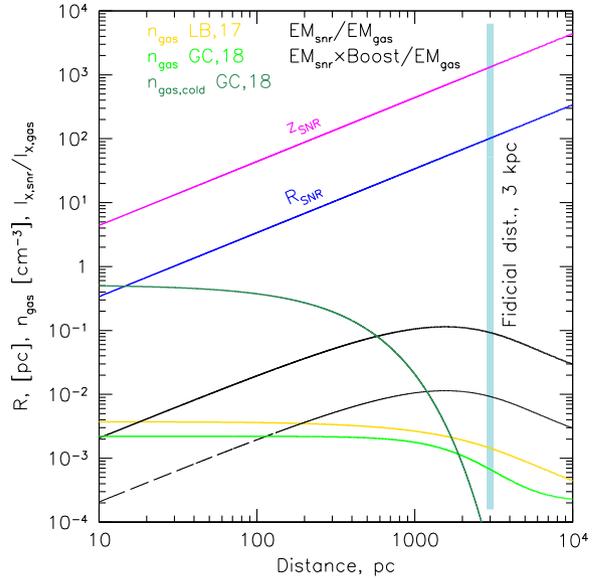}
\caption{Constraints on the SNR position based on the gas density distribution models. The blue and magenta lines show the physical size (radius) of the SNR and its height above the disk plane, respectively,  as a function of the distance from the Sun. The yellow and light-green lines show the hot gas density distributions in the direction of the newly found source according to the models of \protect\cite{2017ApJ...849..105L} 
and \protect\cite{2018MNRAS.474..696G},
respectively. The lower black line shows the ratio of the line-of-sight emission measure ($\int n^2 dl$) of the (hot) gas compressed by the SNR to that of the 
hot MW gas in the same direction. The upper black curve shows the same curve multiplied by the boost factor (10 in this case) due to NEI effects (see \S\ref{sec:nei}. This plot demonstrates that an SNR a few kpc from the Sun could match the most salient properties of the newly found object. An SNR at a distance smaller than $\sim 1-2$~kpc would be a normal SNR in the dense ISM, while the SNR more distant than $\sim 5$~kpc would be too faint compared to the X-ray emission of the Milky Way.
}
\label{fig:halo}
\end{figure}

For our fiducial parameters $R_{max}\sim 200 \,\beta^{-2/3}\,{\rm pc}$. The surface brightness and the total luminosity during the Sedov phase will keep growing until the size reaches $R_{max}$. Therefore, it is more likely to find these very large SNRs in the halo. It is therefore plausible, that the object found in the \texttt{SRG}/eROSITA survey belongs to this class. This is illustrated in Fig.~\ref{fig:halo}, which shows the physical size of the SNR as a function of distance, the (hot) gas density distribution from \cite{2017ApJ...849..105L} and \cite{2018MNRAS.474..696G}. In the latter case, for the flattened (disk-like) component, we used the larger of the two values of the vertical scale-height (namely, $h_z\sim 1.1\,{\rm kpc}$) reported in \cite{2018MNRAS.474..696G} and rescaled the normalization accordingly. For the halo gas, we simply assume that much of the absorbing column density is accumulated over a distance of $\sim 10$~kpc. In addition,  Fig.~\ref{fig:halo} compares the expected surface brightness of the SNR with the background X-ray sky brightness. From this figure, it is clear, that only if the SNR distance is in the range $\sim$2-4 kpc one can explain the appearance of G116.6-26.1 by invoking the NEI effects. Dedicated deep X-ray observations of a few selected fields in the SNR with {\sl XMM-Newton} may help to study plasma equilibrium conditions and to constrain possible shock models.  

As discussed in \S\ref{sec:nei}, spectral information provides constraints on both the upstream ($T_b$) and downstream ($T_f$) electron temperatures at the forward shock. Assuming that the halo gas is in CEI, observations of halo SNRs offer a direct way to measuring the halo gas temperature ($=T_b$). Measuring the gas density of the halo requires the knowledge of the physical size of the SNR (or, equivalently, its distance). One possibility is to use the downstream temperature of electrons derived from the spectra in \S\ref{sec:nei} to estimate the shock velocity. While the ion and electron temperatures might differ strongly when the shock is strong \citep[see, e.g.,][]{2007ApJ...654L..69G}, the assumption that the SNR is in the late stage of its evolution implies that the Mach number is not large and the electron temperature $T_f$ is not  dramatically different from the ion temperature and from its value we can estimate the expansion velocity of the swept gas. More direct measurements of the gas velocity and ions temperature could be provided by future X-ray bolometers with high energy resolution, such as \texttt{XRISM} \citep{2020arXiv200304962X}. The measurement of $v$ would provide a constraint on the density and size via the relation $E\approx \rho R^3 v^2$, where $E$ is known for SNIa. Another constraint should come from X-ray surface brightness $I_X\propto \rho^2 R$. The coefficient of proportionality in this relation is itself a function of other parameters of the problem, but the dependence is not extremely steep (see \S\ref{sec:nei}). Together, these constraints result in estimates of the SNR size and the halo density. We note here that the above arguments are model-dependent. Nevertheless, once the distance is known, we get a new independent way of measuring the density of the hot gas local to the SNR.  High energy resolution would also help measuring the electron temperature on the downstream side of the shock via the comparison of the K$_\alpha$ and K$_\beta$ line fluxes for a given ion. This way, the uncertainties, associated with the departures of the ions fractions from CEI can be avoided. In particular, OVII and OVIII ions seem to be the most promising for this test, given that these ions are the most abundant and there are essentially no Li-like ions of oxygen. 


The supernova remnant detected by the 
\texttt{SRG}/eROSITA appears to be at a stage when the shock wave has already swept about 100 solar masses of the surrounding hot gas, which is dominating the X-ray emission from this SNR. Less clear is the state of matter ejected during the thermonuclear explosion of the white dwarf.  By now, the reverse shock has already swept the ejecta, which consist mainly of iron and other heavy elements with the total mass close to 1.4 solar masses. The outcome of the reverse shock propagation through the iron-dominated ejecta is non-trivial \citep[e.g.][]{1984ApJ...287..282H}. Moreover, in 1D model used here, the ejecta are separated from the shock-heated ISM by the contact discontinuity, while in reality layers of the ejecta and ISM can already be mixed. Mixing affects the gas temperature and efficiency of radiative losses \citep[e.g.][]{2005ApJ...630..864B} posing the question, in which energy band this gas might be detected (if at all). For instance, optical emission in the FeXIV 5303~\AA~ from the freshly shocked ejecta has recently been detected in young Type Ia SNRs \citep{2019PhRvL.123d1101S}. Here we deal with an old SNR and studying in detail the inner regions in different
electromagnetic bands (including X-Rays) would be extremely interesting.

Despite the attractiveness of the "distant SNIa" model, various scenarios involving nearby objects (a few 100 pc) can not be excluded, since a complicated structure of the dust distribution shows some (possibly spurious) signs of correlation with the G116.6-26.1 structure. Also, there are several known high-latitude SNRs, which are apparently embedded in a more dense environment and, therefore, do not belong to the class of "halo SNRs" discussed here \cite[see, e.g.][for recent examples]{2020ApJ...888...90R,2021A&A...648A..30B}. As argued above, future observations should help with the classification of  G116.6-26.1 as a truly distant or more nearby object.   

Finally, we note that even if G116.6-26.1 turns out to be a "regular" SNR, the "type Ia halo SNRs" should exist (see Fig.~\ref{fig:snr_mcmap}). 
If several distant SN~Ia SNRs are identified in the X-ray data (likely with smaller angular sizes), they together will  provide a unique way of probing the halo of the Milky Way at different heliocentric distances. The \textit{SRG}/eROSITA survey, which is designed with the goal of detecting large number of distant galaxy clusters in mind, appears to be particularly well suited for this task.

\section{Conclusions}

A large ($\sim 4^\circ$ in diameter) nearly-circular object SRGe~J0023+3625 = G116.6-26.1 has been found in the data of \textit{SRG}/eROSITA all-sky survey after completion of the first three surveys. The X-ray flux is dominated by emission lines of helium-like (OVII) and hydrogen-like (OVIII) oxygen, with the typical X-ray surface brightness of this emission making $\sim$20\% of the background sky brightness in the 0.5-0.7~keV band.

While the nature of the object is yet to be determined, we argue that many of its features could be explained if it is the remnant of a type Ia supernova occurred in the hot and tenuous gas of the Milky Way halo. If this hypothesis is correct, G116.6-26.1 represents an interesting case of an old ($\sim$40000~yr) and large ($\sim$100~pc) SNR in the gas, which has the cooling time much longer than the SNR age. Departures from the collisional ionization equilibrium downstream of the forward shock produce an order of magnitude jump in the X-ray emissivity, making the source visible in X-rays. As such,  G116.6-26.1 (and similar objects, if more are identified in X-ray data) offers a unique probe of the Milky Way halo temperature, density and metallicity.   

\section*{Acknowledgments}
We thank our referee for useful suggestions that helped to improve the paper and Daichi Tsuna for comments on the failed supernova scenario.

This work is based on observations with the eROSITA telescope onboard \textit{SRG} space observatory. The \textit{SRG} observatory was built by Roskosmos in the interests of the Russian Academy of Sciences represented by its Space Research Institute (IKI) in the framework of the Russian Federal Space Program, with the participation of the Deutsches Zentrum für Luft- und Raumfahrt (DLR). The eROSITA X-ray telescope was built by a consortium of German Institutes led by MPE, and supported by DLR. The SRG spacecraft was designed, built, launched, and is operated by the Lavochkin Association and its subcontractors. The science data are downlinked via the Deep Space Network Antennae in Bear Lakes, Ussurijsk, and Baikonur, funded by Roskosmos. The eROSITA data used in this work were converted to calibrated event lists using the eSASS software system developed by the German eROSITA Consortium and analysed using proprietary data reduction software developed by the Russian eROSITA Consortium.


EC, IK, and  RS  acknowledge  partial  support  by  the  RSF  grant  19-12-00369. AB was supported by the RSF grant 21-72-20020. IIZ acknowledges the support by the IAP RAS state program No. 0030-2021-0005.

\section*{Data availability}
X-ray data analysed in this article were used by permission of the Russian SRG/eROSITA consortium. The data will become publicly available as a part of the corresponding SRG/eROSITA data release along with the appropriate calibration information. 


\bibliographystyle{mnras}
\bibliography{references} 







\bsp	
\label{lastpage}
\end{document}